\documentclass[a4paper]{article}

\usepackage{INTERSPEECH2022}
\usepackage[activate]{microtype}
\usepackage{cite}
\usepackage{color}

\usepackage{comment} 

\title{On the Prediction Network Architecture in RNN-T for ASR}
\name{Dario Albesano\textsuperscript{1}, Jesús Andrés-Ferrer\textsuperscript{1}, Nicola Ferri, Puming Zhan}
\address{Nuance Communications, Inc.}
\email{\{dario.albesano,jesusandres.ferrer\}@nuance.com}
\begin{document}
\maketitle
\begingroup\renewcommand\thefootnote{1}
\footnotetext{Both authors contributed equally to the work.}
\begin{abstract}

RNN-T models have gained popularity in the literature and in commercial systems because of their competitiveness and capability of operating in online streaming mode.
In this work, we conduct an extensive study comparing several prediction network architectures for both monotonic and original RNN-T models.
We compare 4 types of prediction networks based on a common state-of-the-art Conformer encoder and report results obtained on Librispeech and an internal medical conversation data set.
 Our study covers both offline batch-mode and online streaming scenarios.
In contrast to some previous works, our results show that Transformer does not always outperform LSTM when used as prediction network along with Conformer encoder.
Inspired by our scoreboard, we propose a new simple prediction network architecture, $N$-Concat, that outperforms the others in our online streaming benchmark.
Transformer and n-gram reduced architectures perform very similarly yet with some important distinct behaviour in terms of previous context.
Overall we obtained up to 4.1 \% relative WER improvement compared to our LSTM baseline, while reducing prediction network parameters by nearly an order of magnitude (8.4 times).

\end{abstract}

\section{Introduction}
 \label{intro}
 Recurrent Neural Network Transducer (RNN-T)~\cite{RNNTGraves} was proposed as an extension  
 to Connectionist Temporal Classification (CTC)~\cite{CTCGraves}  by relaxing CTC's assumption of conditional label independence between predictions given the audio. 
 In addition to the CTC acoustic encoder, RNN-T introduced a prediction network (PN) and a joiner network.
 The PN autoregressively incorporates previously emitted tokens into the model, while
 the joiner mixes both acoustic and autoregressive label  representations via a monotonic alignment process. 
 Some extensions were proposed to the original RNN-T loss including HAT~\cite{HAT}, RNA~\cite{RNA}, and in particular 
 MonoRNN-T~\cite{MonoRNNT} which constrains the number of output tokens per acoustic input frame in RNN-T alignments.
 %
 Significant architectural improvements have been proposed since the original RNN-T conception.
 LSTM/BLSTM were replaced by Transformer~\cite{TTFacebook,TTGoogle,TTy}, and 
 later  encoder networks were extended with convolutions in the Conformer~\cite{Conformer} and ContextNet~\cite{ContextNet} architectures. 
 In all cases, for streaming recognition the encoder is bound to real-time application requirements.
 These requirements restrict the future context, which leads to an \textit{encoder induced latency} (EIL), that the encoder can exploit.
 Online encoders generally underperform their offline counterparts because of their limited access to future context. Consequently, many approaches 
 were proposed for better leveraging their scarce EIL~\cite{TTGoogle,chunkedMHSA,EMFormer}. 
 Unfortunately, all approaches deemed future context (aka lookahead) as the critical factor. 

 Despite the PN being a core RNN-T differentiator, 
 there is a lack of deep understanding of its role. 
 Some works show that the PN plays a language modeling role~\cite{HAT, ILMT, ILMA} while others
 challenge this interpretation~\cite{Stateless, NonAR, Echo}.
 Under some circumstances, the autoregressive
 dependence of the PN was dramatically simplified to a bigram~\cite{Stateless,TTy,TAR}, and even a randomly initialized PN was not significantly detrimental to RNN-T performance for some tasks~\cite{Echo}.
 Furthermore, there is a full-fledged research line attempting to match RNN-T performance with non-autoregressive CTC-based techniques~\cite{NonAR,pushingCTC}, 
 by dropping the PN and embedding some larger hidden contextual (possibly token) information into the encoder.
 For this reason, the autoregressive context dependence of some specific PN architectures, such as the 
 Transformer-Transducer (T-T) or  the Tied \& Reduced (T\&R) was analyzed~\cite{TTGoogle,TAR}.  %
 For instance, ~\cite{TTGoogle} could achieve similar accuracy by limiting left context to 3 tokens, though observed about 5\% relative degradation when further reducing it to 2 tokens.
 %
N-gram PN 
 was proposed and studied within a causal online Conformer RNN-T encoder~\cite{TAR}.

 This work is devoted to a systematic study of several PNs under various conditions.
 While some partial explorations have been done in the literature, we conduct 
 a clean comparison of several PNs together with a common state-of-the-art Conformer encoder, namely: 
 LSTM~\cite{LSTM}, Transformer~\cite{Transformer}, Conformer~\cite{Conformer}, and n-gram reduced~\cite{TAR}.
 Inspired by~\cite{TAR}, we propose a new n-gram PN that outperforms all the other PNs in online configurations on both the Librispeech 100 hour subset and an internal medical data set consisting of 1000 hours of doctor-patient conversations.
 Motivated by gaining better understanding of the encoder and  PN interactions,
 we assess the PNs in two different regimes, streaming and batch recognition, as well as with character and word-piece vocabularies.
 We further  consider the implications of using either  monotonic or original  RNN-T loss
 within this scoreboard. Finally, we study the behavior of
the best PNs with limited autoregressive context.

\section{RNN-T }

Let $x_1^{T}= (x_1, ..., x_{T})$ be an input sequence of length $T$, where  $x_t\in\mathbb{R}^{d}$ is an acoustic feature vector.   As originally introduced in ~\cite{RNNTGraves}, a RNN-T maps  $x_1^{T}$ to an output token sequence $y_1^U=(y_1, \ldots, y_U)$ of length $U$, where $y_u\in\mathcal{Y}$ is an output token and  $\mathcal{Y}$ is the vocabulary.
This is done via aligning the internal  acoustic and token representations.
We denote an alignment of length $K=T'+U$ as $z_1^K=(z_1, \ldots, z_K)$, where  $z_k$ belongs to  $\mathcal{Y}$ extended 
with a blank symbol $\epsilon$.
$\mathcal{B}^{-1}(y_1^U)$ is the set of all alignments that generate token sequence $y_1^U$ after removing all blank tokens.
Note we have to take into account 3 different lengths, namely, $T$ as original input length, $T'$ as encoder output length, and $U$ as label output length (without blank symbol).
RNN-T models are composed of 3 components: 
%
first \textit{the encoder network}   projects the input acoustics to hidden encoder states 
$e_t=\operatorname{enc}(x_1^{T})$ ($T'<=T$)  for $t=1,\ldots,T'$; 
second  \textit{the prediction network (PN)} autoregressively builds 
a hidden representation of the previous output tokens $s_u=\operatorname{pred}(y_1^u)$ for $u=1,\ldots,U$;
and finally \textit{the joiner network} 
harmonizes both lengths by reading the encoder representation at $t_k$ and the PN representation at $u_k$ to generate the alignment output
distribution, $p(z_k \mid e_{t_k}, s_{u_k})$, over the extended vocabulary.
RNN-T models are optimized to minimize the log-likelihood loss over all possible alignments
\begin{equation}
    \label{eq:rnnt}
    \mathcal{L} = \log p(y_1^U\mid x_1^{T}) 
    = \log \sum_{z_1^K\in\mathcal{B}^{-1}(y_1^U)} p(z_1^K\mid x_1^{T}) 
\end{equation}
with the alignment probability  decomposed left-to-right as
\begin{equation}
    \label{eq:joint_rnnt}
    \textstyle
p(z_1^K\mid x_1^{T}) := \prod_{k=1}^K p(z_k \mid e_{t_k}, s_{u_k})
\end{equation}
where $t_k$ is obtained by counting the number of blanks in the alignment until $k$ and $u_k$ is obtained by counting the number of non-blanks generated until $k$.

As alluded in section~\ref{intro}, two RNN-T variants are analyzed in this work. 
The original RNN-T model~\cite{RNNTGraves} 
allows  alignments with consecutive token sequences without interleaved blanks.
In contrast,  monotonic RNNT~\cite{HAT,MonoRNNT} requires incrementing the encoder position  $t_k$ 
for each token the PN predicts.
In this case, the full alignment length $K$ is $T'$ and $t_k:=k$ in eq.~\eqref{eq:joint_rnnt}.

\subsection{Online Encoder}

Our study of different types of PNs in RNN-T covers both offline batch-mode and online streaming scenario. 
In the online streaming case, the multi-head self-attention (MHSA) and convolutional block in the Conformer encoder 
have to be constrained in accessing future context because of the latency requirement. 
The MHSA block can exploit future context, which is determined by the latency requirement for a specific application, in several ways according to the literature: 
autoregressive~\cite{TAR}, truncated look-ahead~\cite{TTFacebook}, contextual look-ahead~\cite{EMFormer} and  chunked attention~\cite{chunkedMHSA}.
We conducted experiments with these approaches under the baseline LSTM PN and found that the chunked attention~\cite{chunkedMHSA} provides the best trade-off (see section 3.1) between latency and accuracy. Therefore, we use this  approach, along with causal depth-wise convolutions in the convolutional block, to evaluate different PNs in the online scenario.

\subsection{Prediction Networks (PN)}

In this work, we mainly compare 4 standard PNs: LSTM, Transformer, Conformer, and n-gram.
The LSTM PN consists of one layer unidirectional LSTM~\cite{LSTM} to  model the full left context,
although some research has exploited limited left context ~\cite{LessIsMore}.
For the Transformer PN, the MHSA is autoregressively masked with limited left context.
For the Conformer PN~\cite{Conformer}, causal convolutions and 
MHSA blocks were used in a similar way as in the Transformer case. 
In the next subsection, we describe n-gram PN, specifically the N-Avg~\cite{TAR} 
as well as our proposed  N-Concat extension in more detail.

\subsubsection{Reduced n-gram encoder}
\label{sec:ngram}
%
\textbf{N-Avg :} Tied and reduced PN~\cite{TAR} simplifies the autoregressive dependency to an n-gram dependency.
 The previous $N-1$ tokens are used to generate the $s_{u_k}$ embedding representation.
 First, a shared embedding matrix projects each previous token $y_{u-n}$ into a $D$-dimensional embedding vector $v_{u-n}$. 
 Next,  the embedding vectors,  $v_{u-n}$, are scaled by the dot product between themselves
 and the  positional encodings, $q_{n,h}$, for each head; and then averaged across the full $N$ left context positions and $H$ heads.
 Finally, a  feed-forward neural-network is applied to the averaged representation.
 Specifically, the PN state $s_u$ at $u$, which is used to predict $y_{u+1}$,
 is computed as 
 \begin{equation}
     \label{eq:tar}
     s_u := \operatorname{LayerNorm}( \operatorname{Projection}(s_u^{(\text{avg})}))
 \end{equation}
 with
  \begin{equation}
     \label{eq:tar_avg}
     \textstyle
     s_u^{(\text{avg})} := \frac{1}{H} \sum_{h=1}^{H} \frac{1}{N} \sum_{n=0}^{N-2} [v_{u-n}\odot q_{h,n} ] \cdot v_{u-n}
 \end{equation}
where $\odot$ is the dot product and $\cdot$ is a scalar-to-vector  multiplication.

\textbf{N-Concat:} N-Avg averages along heads and the context tokens.
In order to promote an explicit bias towards attending  different context tokens,  we propose  a new variant, N-Concat, that incorporates
concatenation heads instead of averaging heads, analogous to standard MHSA~\cite{Transformer}.
This newly proposed PN allows retention of compressed information from each context token in the final representation.
This PN splits the embedding dimension $D$ into $H$ vectors along which it applies a single averaging head from N-Avg.
Let $s_u(m)$  denote the $m$-th split of the embedding dimension
, N-Concat computes
   \begin{equation}
     \label{eq:tar_concat}
     \textstyle
     s_{u}^{(\text{concat})}(m) := \frac{1}{N} \sum_{n=0}^{N-2} [v_{u-n}(m)\odot q_{n}(m) ]\cdot v_{u-n} (m)
 \end{equation}
 and then concatenates each into the pre-projected representation
 \begin{equation}
 \textstyle
  s_{u}^{(\text{concat})} = (s_{u}^{(\text{concat})}(1), s_{u}^{(\text{concat})}(2), \ldots, s_{u}^{(\text{concat})}(D/H))
 \end{equation}
 Similar to  N-Avg,  a LayerNorm and Projection is applied to the concatenated representation, $s_u^{(\text{concat})}$.

\section{Experiments}
\label{sec:exps}
\begin{table}[t]
  \caption{Baseline WER results with Conformer encoder and LSTM PN with various online MHSA configurations for character-based LS100. LA stands for lookahead.}
  \label{tbl:ls100-online-mods}
  \centering
  \begin{tabular}{lccc}
    \toprule
     Encoder &  EIL & test-clean & test-other \\ 
    \midrule
      Ours Offline                         & n.a. &  $\mathbf{5.9}$  & $\mathbf{16.9}$ \\
    \phantom{chk} + causal convs           & n.a. &  $6.0$  & $17.3$ \\  
    \midrule
       auto-regressive                      & 0.04 &  $9.2$  & $25.4$ \\
       \phantom{chk}  + 0.96s LA            & 1.00 &  $7.6$  & $22.6$ \\
       chunked (0.5s)                       & 0.52 &  $7.6$  & $22.3$ \\
       \phantom{chk} + 0.5s LA              & 1.00 &  $7.1$  & $21.5$ \\
       \phantom{chk} + 0.5s contextual  LA  & 1.00 & $\mathbf{6.8}$  & $\mathbf{20.4}$ \\ 
       chunked (1s)                         & 1.02 &  $7.3$  & $21.5$ \\
    \bottomrule
  \end{tabular}
\end{table}

We conducted our experiments on two datasets, Librispeech and an internal medical dataset, using different PNs.
In both cases, we extracted 80-channel filterbank features computed from a 25ms window with a stride of 10ms,
followed by SpecAugment~\cite{SpecAugment} (without time warping)
and finally a small front-end consisting of 2 convolutional layers that perform downsampling by a factor of 4.
In all cases, we used Adam~\cite{ADAM} and warming-up learning rate schedule~\cite{Transformer} tuned as described below.
Similarly, we kept the same  Conformer encoder architecture~\cite{Conformer} while varying the PN.


For the Librispeech corpus~\cite{Panayotov2015-LAA},  we selected the 100 hour training subset (denoted as LS100), 
and mitigated overfitting by augmenting the data by a factor of 3 
with speed perturbation~\cite{Ko2015-AAF} factors in the range of \{0.9, 1.0, 1.1\}.
We used 2 different output token sets:  30 characters and 300 word-pieces with an average length of $4.78$ characters.
All models were trained for 100 epochs on the augmented LS100 data set.
The training hyperparameters were tuned based on the offline model with Conformer encoder and an LSTM PN using a 
limited grid search on the clean development set of LS100, then applied to all other models.
For assessing the PNs, we used an encoder of 18 Conformer blocks with hidden size of 256, feed-forward (FF) dimension 1024,
convolutional kernel width of 15 and 4 attention heads, followed by a joint network of two 256-dimension FF layers with ReLu activation.
For the PN, we used a baseline LSTM PN of 256 hidden units; a Transformer PN with 256 hidden dimension, 1024 FF dimension, and 4 heads;
a Conformer PN with the same size as the transformer PN plus a convolutional kernel width of 15 tokens; and finally for both n-gram PNs (see section~\ref{sec:ngram}) 
we used 4  heads  followed by a dense layer of 256. 
All PNs have a single layer since we found them to outperform deeper ones in preliminary experiments,
in contrast to some works~\cite{TTFacebook, TTGoogle, TTy, NonAR}.

\begin{table}[t]
  \caption{WER on LS100 varying the PN in offline RNN-T models }
  \label{tbl:ls100-offline}
  \centering
  \begin{tabular}{lccc}
    \toprule
     PN &  output units & test-clean & test-other \\
    \midrule
      LSTM~\cite{NonAR} & word-pieces    &  $6.8$  & $18.9$ \\
      LSTM (ours)         & word-pieces & $6.1$  & $16.9$ \\
  \midrule
      LSTM          & characters & $5.9$  & $16.9$ \\
     Transformer    & characters &  $\mathbf{5.6}$  & $16.9$ \\
     Conformer      & characters & $5.7$  & $\mathbf{16.4}$ \\ 
     N-Avg        & characters & $\mathbf{5.6}$  & $16.5$ \\
     N-Concat     & characters& $5.7$  & $16.9$ \\
    \bottomrule
  \end{tabular}
\end{table}

Additionally, we studied different configurations on an internal 
medical speech transcription task of doctor-patient conversations (\textit{D2P1K}), 
across multiple medical specialties. 
We use a lab setup where models were trained on 1k hours with a 2.5k word-piece vocabulary 
of $6.2$  average character length.
Word error rate (WER) was computed on a speaker-independent test set consisting of 263k words in total~\footnote{
 WER numbers for both datasets were rounded so that differences remain significant statistically.
}.
We tuned the hyperparameters based on the offline LSTM PN baseline, then applied them to the proposed configurations without further tuning.
In this case, we used larger and shallower Conformer encoder of 16 blocks with hidden size of 512, 
 FF dimension of 1024, 8 heads and convolutional kernel width of 17
 followed by the same joint network as in LS100.
Similarly we used larger PN: 
the LSTM PN baseline is enlarged to 640 hidden units; 
the Transformer PN is expanded to  640 hidden dimension, 1024 FF dimension, and 8 heads;
 the Conformer PN uses on top of the transformer PN a convolutional kernel width of 17 tokens;
and finally  in both n-gram PNs  the dense layer is increased to 640 units.
We also observed  that a single layer PN performed the best in this case.
For instance, a 2 layer Transformer PN  degraded by 3 \% relative WER; and a $2$ layer  N-concat by 1\%. 

\subsection{Results With Different Online Approaches in Encoder}

Table~\ref{tbl:ls100-online-mods} reports the results with the different online encoder 
approaches described in section 2.1 with  LSTM PN. The contextual look-ahead encoder 
gave the best accuracy, but at the expense of twice inference time.
Conversely, 
the 1 sec chunked attention encoder offered a competitive trade-off.  
Based on these results, we use the 1 sec chunked attention in the Conformer encoder in the following study.
Note that online models degrade more significantly
under adverse acoustic conditions and/or domain mismatch as showed in the results on test-other set.

\begin{table}[t]
  \caption{%
  WER on LS100 tests sets for online models with 1 second chunked encoder trained with two output units:
characters and word-pieces. PN parameters are reported as \textit{\# params}.
           }
  \label{tbl:ls100-online}
  \centering
  \begin{tabular}{lccccccc}
    \toprule
     Output label & & \multicolumn{2}{c} { characters }     
                  & \multicolumn{2}{c}{ word-pieces } \\
           \midrule
     PN     & \# params & clean & other &  clean & other \\
      \midrule
     LSTM          & $0.59$M &  $7.3$          & $\mathbf{21.5}$ & $7.3$ & $21.5$ \\
     Transformer       & $0.9$M  &  $\mathbf{7.1}$ & $22.1$ & $7.4$ & $22.0$ \\
     Conformer         & $1.6$M  &  $7.3$  & $21.8$& $7.5$ & $21.9$ \\
     N-Avg       & $0.09$M & $\mathbf{7.1}$  & $21.9$ & $7.3$ & $21.3$ \\
     N-Concat    & $0.07$M &  $\mathbf{7.1}$ & $21.9$ & $\mathbf{7.0}$ & $\mathbf{21.2}$ \\
    \bottomrule
  \end{tabular}
\end{table}

\subsection{Results With Different Prediction Networks (PNs)}

Table~\ref{tbl:ls100-offline} 
reports offline results with different PNs 
on the LS100 dataset with characters as outputs.
The results show that the N-Concat performs slightly worse than N-Avg, but better than the LSTM, and competitive with the Transformer and Conformer PN.
We also include the corresponding results from ~\cite{NonAR} and our LSTM PN result with word-piece units in the table, because our Conformer encoder and LSTM PN have similar configurations as theirs. 
Note our baseline model is competitive with theirs.

Table~\ref{tbl:ls100-online}
reports results from different PNs based on the online Conformer encoder with 1 second chunked-attention. Results are on the LS100 set with both character and word-piece units. 
For all PN except LSTM, we limit the left context 
to $5$ word-pieces or $24$ characters, so that a similar amount of  context
(the average word-piece length is $4.78$ characters) is accessed in both cases.
It shows a different picture to that of offline encoders in Table~\ref{tbl:ls100-offline}.
N-Concat works best across the board, while the Conformer PN increases the number of parameters  without providing any improvement.
The LSTM PN is competitive on test-other with character units, but lags behind on test-clean.
The Transformer PN is competitive only on test-clean and only with character units, probably due to the large number of parameters.
Both N-Avg and N-Concat  perform the same with character units, but N-Concat outperforms N-Avg by  4.1\% relative WER reduction with word-pieces while also reducing the number of parameters by 22.5\% 
(or $8.4$ times with respect to the LSTM PN).

 %
 In Table~\ref{tbl:dax},  different PNs are evaluated on the D2P1K task.
 Overall, a similar picture to that of LS100 is observed.
 %
 While the LSTM PN is competitive  in offline case,
 it clearly lags behind in the online case where the performance of different PNs vary significantly. 
 In the online case, N-Concat performs best, though not far from N-Avg and Transformer PN.
 We hypothesize that the encoder attempts to model the entire transduction task by itself and 
 the  PN mainly complements it for disambiguation when conditions become adverse.
 At the same time, the token prediction task is perhaps learnt faster and consequently 
 a strong model with a long left context dependency is detrimental for generalization
 because of sparse statistics.
 
\begin{table}[t]
  \caption{WER for different PNs 
         in both offline and online encoder regimes 
         for  RNN-T models on D2P1K corpus}
  \label{tbl:dax}
  \centering
  \begin{tabular}{lccc}
    \toprule
     PN &  Num. PN params &\multicolumn{2}{c} { WER } \\
      &   & Offline & Online \\
      \midrule
          LSTM      & $3.69$  M &  $13.2$  & $14.6$ \\
     Transformer    & $3.78$  M &  $\mathbf{12.9} $ & $13.9$ \\
     Conformer      & $5.10$  M &  $\mathbf{12.9} $ & $14.0$ \\
     N-Avg        & $0.42 $ M &  $\mathbf{12.9} $ & $13.9$ \\
     N-Concat     & $0.41 $ M &  $13.0 $ & $\mathbf{13.8}$ \\
    \bottomrule
  \end{tabular}
 \end{table}

 \begin{table}[t]
  \caption{WER and RTF with different PNs for online Monotonic RNN-T on D2P1K. Arrows point to improvement direction. Benchmarked on V100 GPU. 
  }
     \label{tbl:monoRNNT}
  \centering
    \begin{tabular}{rccccc}
    \toprule
       & LSTM & Transf. & Conf. & N-Avg & N-Concat \\
     \midrule
     WER$\downarrow$ &  $14.2$  & $\mathbf{14.1}$  & $\mathbf{14.1}$ & $14.3$ & $14.2$ \\
     RTF$\downarrow$ &  $1.22$  & $1.18$           &  $1.23$         & $1.18$ & $\mathbf{1.17}$ \\
     BS$\uparrow$    &  $2.5$K  &  $2.4$K          &  $2.2$K         & $2.5$K & $\mathbf{2.6}$K \\
     \bottomrule
  \end{tabular}
\end{table}

  Finally, in Table~\ref{tbl:monoRNNT}, we assess the effects of adding the monoRNN-T constraint to the loss in the online  setup. We also report the real time factor (RTF) measured with 3K utterances as batch size  in greedy decoding and the maximum batch size  at 1 RTF (\textit{BS} row). Note that the RTF numbers are measured via forward propagating through the whole network. Since the encoder, which contains 64M parameters, dominates in size, the overall RTFs look similar, despite significant reduction  in N-Avg and N-Concat PN size. We use 5-gram left context for all but LSTM. 
  Accuracy slightly degrades with respect to Table~\ref{tbl:dax}. However, the relative differences among all the PNs are very small.
  It can be seen that N-Concat is the best one in terms of RTF.

\subsection{Reducing Left Context in Prediction Networks}

  Figure~\ref{fig:left} shows the curves of WER under different left context length for Transformer and N-Concat PN.
  The graph on top is from varying left context length only at inference based on a model trained with 128 left contexts.
  The graph at the bottom left is from varying left context length at both training and inference (i.e. matched left context) with word-piece units. The bottom right graph is the same as the left one but with character units.
  %
  We also conducted multiple similar experiments as the top graph in Fig.~\ref{fig:left}), but based on models trained with different left context length (i.e. not 128 as showed in Fig.~\ref{fig:left}).
  Our finding is that, as showed in the top graph, the N-Concat PN reaches maximal performance when using
  4 left tokens at inference regardless of the size of left context used in training and it is more resilient to the changes in left context length.
  In contrast, for the Transformer PN, matched left context length at training and inference is needed for achieving maximal accuracy, though it still underperforms the N-Concat PN in such case and it is much more sensitive in short left context length conditions.
  %
  When varying left context length in a matched way at both training and inference, a similar picture is observed, where N-Concat improves over Transformer PN aided by the regularization effect of having less parameters.
  Considering that the average utterance length in the training set is $78$ word-pieces ($186$ characters), the Transformer PN seems not be able to exploit long-term memory at training either.
  %
  This is particularly observed on the test-other test and aligned with the regularization effect that limited context has on both models.
  Stateless or single token context hurts performance the most,
  especially for character models, for which we used early stopping to prevent divergence.
  We observed similar though more consistent trends on the D2P1K set that we omit for lack of space.
  However, in Table~\ref{tbl:leftDax}  we report the most representative data points  during training for this task.
 
\begin{figure}[t]
 \centering
 \includegraphics[width=1.0\linewidth]{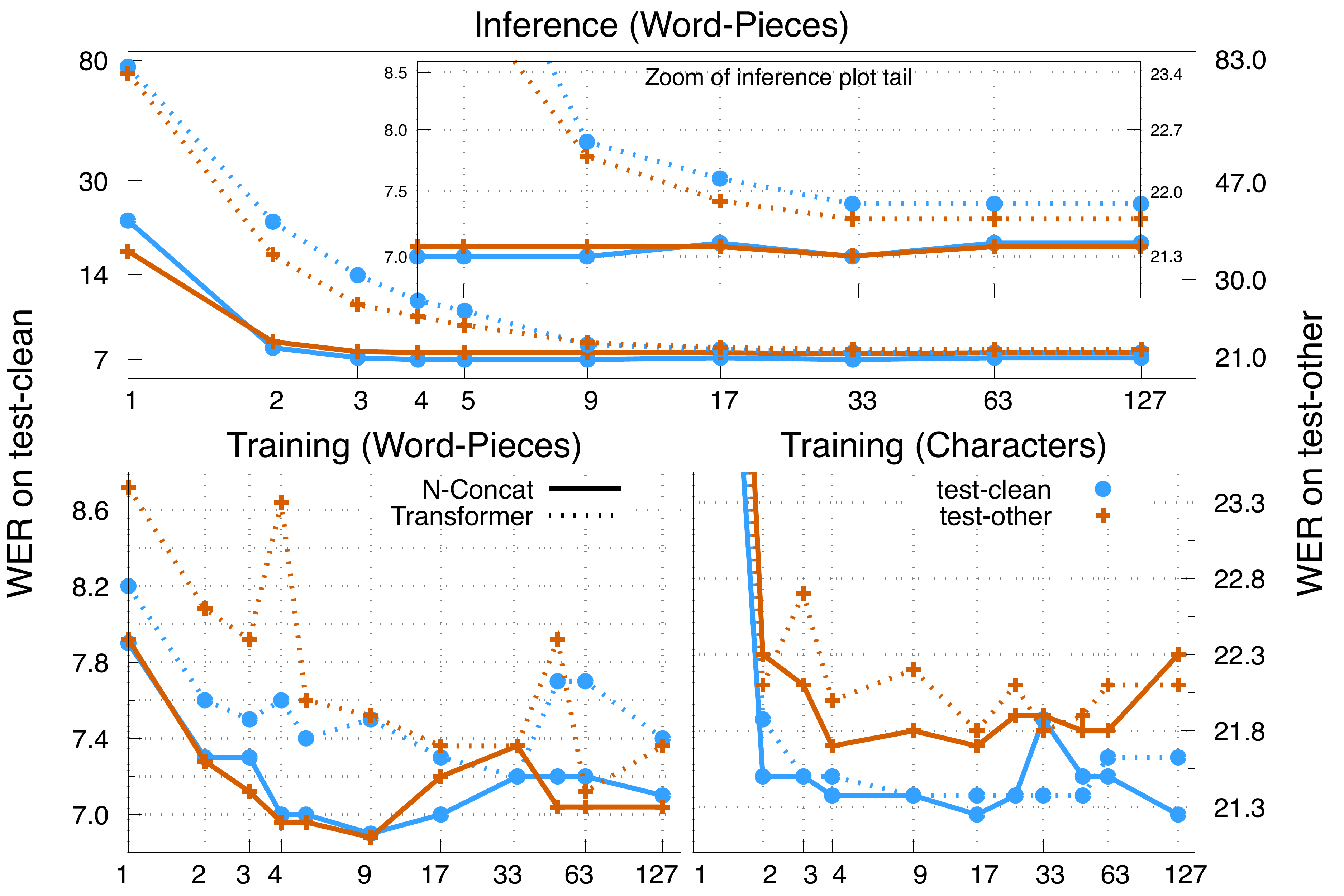}
 \caption{
 $N$-Concat and Transformer WER on LS100 as a function of the left context length.
  Both test-clean (blue line) on the left y-axis and test-other (brown dashed line) on the right secondary y-axis are shown in each plot. Legends are shared across all plots. All models use 1 sec EIL chunked encoder.
 }
 \label{fig:left}
 \end{figure}

\begin{table}[t]
  \caption{ Effect of limiting  left context 
  for  Transformer (as Transf.) and $N$-Concat PNs on D2P1K dataset measured in WER.} 
  \label{tbl:leftDax}
  \centering
  \begin{tabular}{ cccccccc}
    \toprule
      left-context         & 1 & 2 & 3 & 4 & 5 & 6  \\
     \midrule
      Transf.          &   $14.6$  & $14.0  $  & $13.9  $  & $13.9 $  & $14.0 $  & $14.0 $ \\
       $N$-Concat     &   $14.4 $  & $14.0$  & $13.9$  & $13.9$  & $\mathbf{13.8}$  & $14.0$ \\
    \bottomrule
  \end{tabular}
\end{table}

\begin{table}[htb]
  \caption{ WER on test-clean LS100 as we force $2$-Concat online RNN-T to behave like a unigram with a regularization weight.}
  \label{tbl:1gram}
  \centering
  \begin{tabular}{ccccccc}
    \toprule
      Reg. weight  & 0.0 & 0.01 & 0.1 & 1 & 10  & 25 \\
     \midrule
      $2$-Concat &  $7.9$  & $ 8.0$  & $7.6$  & $ 7.9 $  & $9.7$ & $24.3$\\  
    \bottomrule
  \end{tabular}
 \end{table}

 Table~\ref{tbl:1gram} fills the gap between a 1-gram PN (or non-autoregressive), which always diverged in our case,
 and a bigram word-piece PN by adding a l2 regularization that forces each embedding to be as close as possible to the average learnt embedding.
 The larger the regularization weight, 
 the more non-autoregressive the model is. 
 As expected, the performance tends to the grade as the regularization weight is increased.

\section{Conclusions}

  We proposed a N-Concat PN that reduces the PN parameters  while still outperforming the other PNs in online RNN-T models. 
  We ran extensive experiments with 
  different PNs and found that their performance heavily depends on the acoustic
  conditions and the encoder regime, either online or batch recognition.
  In online scenarios the PN becomes more important and  an overparametrized PN can hurt performance.
  For offline recognition, we found that the PN architecture becomes less important than in  the online case.
  Although both Transformer and N-Concat PN are resilient to limited left context at training and inference,
  N-Concat more efficiently exploits the left context.
  Both PNs are not able to exploit long-term dependencies and yet limiting the left context to less than 2 tokens
  severely degrades performance.
  %
   %
   As future work, we would extend our study to take external language models into account, 
   and exploit the limited n-gram dependency  to generate lattices as outputs for downstream tasks.
\clearpage

\bibliographystyle{IEEEtran}

\bibliography{mybib}

\end{document}